\documentclass[10pt,preprint]{emulateapj}

\newcommand{\cbo}{L429}

\newcommand \msun{\hbox{$\hbox{M}_{\odot}$}}

\begin{document}

\title{\textit{Spitzer} Observations of L429: A Near--collapse or
  Collapsing Starless Core}

\author{Amelia M. Stutz\altaffilmark{1},
  Tyler L. Bourke\altaffilmark{2},
  George H. Rieke\altaffilmark{1},
  John H. Bieging\altaffilmark{1},
  Karl A. Misselt\altaffilmark{1},
  Philip C. Myers\altaffilmark{2},
  Yancy L. Shirley\altaffilmark{1}}

\altaffiltext{1}{Steward Observatory, University of Arizona, 933 North
  Cherry Avenue, Tucson, Arizona 85721; astutz@as.arizona.edu.}

\altaffiltext{2}{Harvard-Smithsonian Center for Astrophysics, 60
Garden Street, Cambridge, MA 02138}

\begin{abstract}

We present {\it Spitzer} infrared observations of the starless
core \cbo.  The IR images of this core show an absorption feature,
caused by the dense core material, at wavelengths~$\leq$~70~\micron.
The core has a steep density profile, and reaches $A_V > 35$~mag
near the center.  We show that \cbo\ is either collapsing or in a
near--collapse state.

\end{abstract}

\keywords{ISM: globules -- ISM: individual (LDN 429) -- infrared: ISM
  -- (ISM:) dust, extinction}

\section{Introduction}

The collapse of a dense cloud core into a star happens rapidly
\citep{hayashi66} and therefore is rarely observed; furthermore, most
dense cores appear not to be collapsing but to be close to equilibrium 
\citep[e.g.,][]{lada08}.  If they are not sufficiently supported by
gas pressure and turbulence, then a plausible level of magnetic field
strength would suffice to support them
\citep[e.g.,][]{kandori05,stutz07,alves08}.  Because the process of
collapse is so rarely observed, there is considerable debate about
what triggers it and how it proceeds to form a star.  Additional
examples of dense cores in extreme states that indicate either
collapse or a condition just prior to collapse are critical to
constrain theories about the first stages of star formation.

Mm-wave line observations are the traditional means to search for
collapsing cloud cores, such as in the pioneering study by
\citet{walker86}. The subsequent discussion
\citep[e.g.,][]{menten87,walker88,wooten89,mundy90,narayanan98}
illustrates the ambiguities in the interpretation of velocity data in
star forming dense cores.  Alternative observational approaches are
necessary to supplement such line measurements and provide a better
understanding of collapsing and nearly-collapsing dense cores.

In this paper we characterize the structure of the extremely dense
core \cbo, using the infrared shadow technique we have demonstrated on
two other globules \citep{stutz07,stutz08}.  This core is already
known to have an extremely high deuterium fraction, believed to be an
indication of its being very close to collapsing into a protostar, and
a low internal temperature of only $T_{\rm kin}\!\sim\!8$~K, reducing
pressure support to low levels \citep{bacmann03,crapsi05}.
\citet{bacmann00} detected a 7~\micron\ shadow from \cbo\ using
ISOCAM.  Here, we report that this shadow is so opaque that it is
detectable at 70~\micron.  The presence of significant absorption at
this long wavelength demonstrates that \cbo\ is indeed extremely
dense.  Our analysis of the shadow also demonstrates that it is very
compact with a steep density gradient.  We show how these
characteristics put \cbo\ in the collapsing or near--collapsing
starless core category.

\section{Observations and processing}

\begin{figure*}[t]
   \begin{center}
     \scalebox{0.772}{{\includegraphics{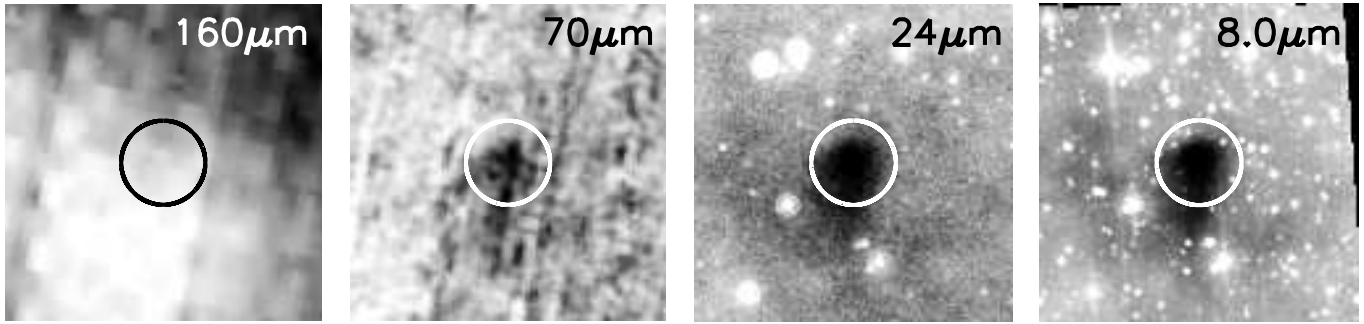}}}
     \caption{MIPS and IRAC $5\arcmin \times 5\arcmin$ images of \cbo\
       centered at RA = $18^h 17^m 05.4^s$, Dec = $-08\degr 13\arcmin
       32.3\arcsec$ (J2000; $l = 21.68\degr$, $b = 3.80\degr$).  The
       circle (radius $= 40\arcsec$) indicates the location of the
       shadow produced by the dense core material.  The images are
       displayed on a log scale; north is up and east is to the left.}
     \label{fig:img}
   \end{center}
\end{figure*}

\cbo\ was observed by {\it Spitzer} using MIPS, program ID 30384,
P.I.\ Tyler L. Bourke, and with the IRAC instrument, program ID 20386,
P.I.\ Philip C. Myers.  The MIPS observations utilized scan map mode
in all three bands.  The data reductions were carried out as described
in \citet{stutz07}.  The MIPS images are $\sim\!15\farcm5 \times
54\arcmin$ in size, with exposure times of $\sim\!150$~s at
24~\micron, $\sim\!60$~s at 70~\micron, and $\sim\!15$~s at
160~\micron.  For the 70~\micron\ and 24~\micron\ data respectively
the beam size is $\sim\!18\arcsec$ and $\sim\!6\arcsec$, the nominal
default pixel scale is $9\farcs8$ and $2\farcs5$, and the pixel scale
in our images is $4\arcsec$ and $1\farcs24$.  The IRAC reductions were
carried out as described in \citet{harvey06}.  The IRAC images are
$\sim\!6\arcmin \times 5\farcm5$ in size, with exposure times of
$\sim800$~s.  The {\it Spitzer} data are shown in
Figure~\ref{fig:img}.  We also use the 2MASS All--Sky Point Source
Catalog (PSC) \citep{skrutskie06} photometry for two well--detected
sources within $40\arcsec$ of the center of \cbo
Reddening estimates for these sources, assuming the \citet{rieke85}
extinction levels, are plotted in Figure~\ref{fig:av}.

\section{The Distance}

The commonly accepted distance to \cbo\ is 200~pc
\citep[e.g.,][]{lee99,bacmann00,crapsi05}, based on the likely
association of this cloud with the Aquila Rift.  \citet{dame85} find a
distance to the Aquila Rift based on the CO line V$_{\rm LSR} =
8$~km~s$^{-1}$.  They summarize the results of various studies and
conclude that the distance to the Aquila Rift is consistent with
$200\pm100$~pc.  More recently, \citet{straivys03} derive a distance
to the near side of the Rift of $225\pm55$~pc, while
\citet{kawamura01} derive a distance to an Aquila cloud complex
centered at $\sim\!l = 48\degr$ and $\sim\!b = -6\degr$ of
$230\pm30$~pc, both consistent with previous studies.  The depth of
the absorption feature (see Fig.~\ref{fig:img}) implies that \cbo\
must lie on the near side of, or in front of, the Aquila Rift because
substantial amounts of foreground material would be warmed by the
interstellar radiation field and would dilute the shadow.  We assume a
distance of 200~pc, consistent with previous distance determinations.
If the distance to \cbo\ is larger the derived mass will increase,
making the core less stable and more likely to be collapsing.

\section{Shadow Profiles}

\cbo\ is seen in absorption at $\lambda \leq$~70~\micron; at
160~\micron\ there is large-scale emission but no isolated local peak
is observed (see Fig.~\ref{fig:img}).  In addition, no compact
emission source is detected at any {\it Spitzer} wavelength within the
shadow, and so no source of luminosity $\gtrsim 10^{-3}$~L$_{\odot}$
is present \citep{harvey06,dunham08}.  We derive extinction profiles
from the observed shadows at 70, 25, 8.0, and 5.8~\micron\ using
techniques similar to those discussed in detail in
\citet{stutz07,stutz08}.

\subsection{70~\micron\ profile}

The first step in deriving the 70~\micron\ profile is to subtract from
the image a value that represents zodiacal emission and other
foreground sources.  For \cbo\ the shadow is the darkest part of the
image.  For this reason the image zero level cannot be determined
directly from the image itself.  To set an upper limit on the optical
depth, we zero the image using the lowest pixel value in a $44\arcsec
\times 44\arcsec$ box centered on the shadow; in this case the zero
value is $\sim\!5.2$~MJy~sr$^{-1}$ below the mean pixel flux level
immediately outside the shadow.  Ideally, one would compute the
minimum value based on more pixels.  However, the 70~\micron\ pixel
size is a non-negligible fraction of the size of the core of
absorption feature; taking more pixels would dilute the depth of the
shadow.  We use IRAS data to set the lower limit on the optical depth
profile: we measure the flux level outside the shadow to be
$\sim\!13.5$~MJy~sr$^{-1}$ above the foreground component of emission,
interpolated to 70~\micron\ from the 60 and 100~\micron\ IRAS images
and estimated far away from regions with significant emission (about
$4.5\degr$ away).  For this value, the minimum shadow value is
$8.3$~MJy~sr$^{-1}$ above the image zero level.  This zero level is a
very conservative estimate because the shadow region is surrounded by
high levels of emission which are very likely to raise the foreground
contribution above the level we use here.  This lower limit also
accounts for the MIPS 70~\micron\ calibration error, $5$\%
\citep{gordon07}, and the IRAS extended source calibration error,
which can be as large as $30$\% \citep{wheelock94}.

\begin{figure}
  \begin{center}
     \scalebox{0.46}{{\includegraphics{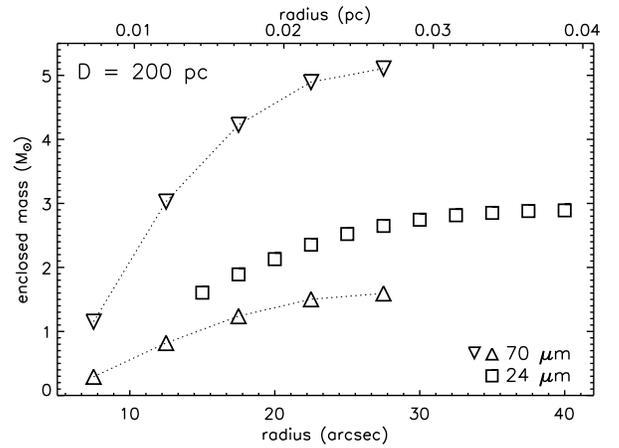}}}
     \caption{Enclosed mass vs.\ radius for \cbo\ calculated assuming
       a distance of $200$~pc.  The downward-- and upward--pointing
       triangles represent masses calculated using the upper and lower
       bounds on the 70~\micron\ extinction profiles, respectively.
       The 24~\micron\ masses (squares) are scaled to match the
       70~\micron\ profile; we only show 24~\micron\ points with $\tau
       < 1.0$.  We assume \citet{ossenkopf94} dust opacities (see \S
       4).}
     \label{fig:mass}
  \end{center}   
\end{figure}

After zeroing the image we mask out all pixels with possible emission
above the sky level.  The mask threshold is set to the mean value plus
$0.5\sigma$ measured for pixels in a $4\arcmin \times 4\arcmin$ region
$4\arcmin$ north of the shadow.  Next we measure the mean pixel
values, termed $f$ here, in nested, and concentric, circular regions,
with radii starting at $7\farcs5$ and increasing in $5\arcsec$ steps,
out to $\sim\!120\arcsec$, centered at RA = $18^h 17^m 05.4^s$, Dec =
$-08\degr 13\arcmin 32.3\arcsec$ (J2000), without accounting for
fractions of pixels.  We ignore all pixels in the southern half of the
shadow because the shadow has a tail extending to the south and the
assumption of spherical symmetry breaks down (see
Figure~\ref{fig:img}).  Finally, to determine the optical depth at
70~\micron, $\tau_{70} = -ln(f/f_0)$, we average profile values at
radii $> 27\farcs5$ to measure $f_0$, the intrinsic, unobscured local
flux value.  The outer radius of the core at 70~\micron, equal to
$27.5\arcsec$, is set by the observed flattening of the measured flux
profile at that radius, indicating that there is little to no
absorption at larger radii.  Following \citet{stutz07}, we derive the
70~\micron\ $A_V$ profile shown in Figure~\ref{fig:av} with open
triangles: downward--pointing triangles represent the upper limit and
the upward--pointing triangles represent the lower limit.  The
innermost region has an optical depth of $\tau_{70} = 1.8$ to $0.46$
(upper and lower limits).  The total mass (gas plus dust) implied by
the 70~\micron\ profile is given by
\begin{equation}
M_{70} = 5.1~M_{\odot} \left( \frac{D} {200~{\rm pc}} \right)^2
\frac{1.58 \times 10^2~{\rm cm}^2~{\rm gm}^{-1}} {\kappa_{\rm
abs,70}},
\end{equation}
where we have assumed a gas to dust mass ratio of 100 and the
\citet{ossenkopf94} model dust opacity for coagulated dust grains of
$1.58 \times 10^2$~cm$^2$~gm$^{-1}$.  The value of $5.1$~\msun\ is the
{\it upper limit} on the mass; the {\it lower limit} mass is
$1.6$~\msun.  The enclosed mass profile is shown in
Fig.~\ref{fig:mass}.  As noted earlier, our limits on $\tau$ are very
conservative.  We performed an error analysis to measure the effect of
a reasonable perturbation in the assumed zero level and the calculated
value for $f_0$ and find that the errors are absorbed into the limits
by a wide margin.  For a dust opacity of $\kappa_{\rm abs,70} = 6.35
\times 10^{1}$~cm$^2$~gm$^{-1}$, from the $R_V = 5.5$ model from
\citet{draine03a,draine03b}, we calculate a 70~\micron\ mass of
$M_{70} = 12.6 - 4.0$~\msun.

\subsection{24, 8.0, and 5.8~\micron\ profiles}

\begin{figure}
  \begin{center}
     \scalebox{0.46}{{\includegraphics{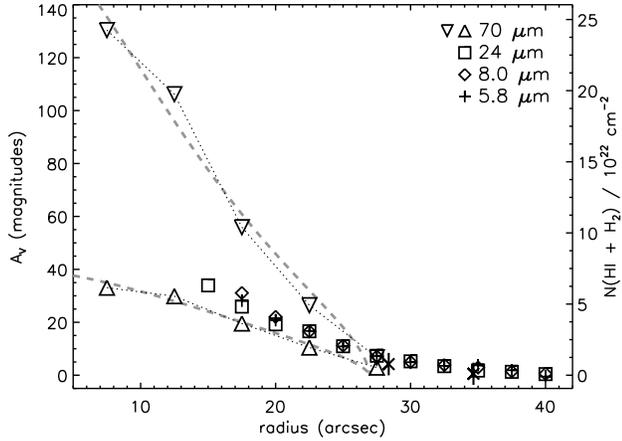}}}
     \caption{Extinction profiles for \cbo.  The downward-- and
       upward--pointing triangles (connected with a dotted line)
       represent upper and lower bounds on the 70~\micron\ extinction
       profile, respectively; the innermost 70~\micron\ region has an
       optical depth of $\tau_{70} = 1.8$ to $0.5$, for the upper
       and lower bounds, respectively.  The dashed lines represent the
       Bonnor-Ebert best-fit models for the 70~\micron\ extinction
       profile derived assuming a distance of 200~pc; both models
       require high temperatures that are in contradiction with the
       observations (see \S~5).  The 24~\micron\ (squares),
       8.0~\micron\ (diamonds), and 5.8~\micron\ (plus signs) profiles
       are scaled to match the outer two radial regions of the
       70~\micron\ profile; for these three profiles we only show
       points with $\tau < 1.0$.  The crosses ($\times$) are 2MASS
       extinction measurements for two sources with reliable
       detections.  We assume \citet{ossenkopf94} dust opacities for
       the extinction profiles (see \S~4).}
     \label{fig:av}
  \end{center}   
\end{figure}

The method used to derive the shorter wavelength profiles is
essentially the same as described above.  First, we zero the images
using the darkest pixel in the shadow.  Second, the mask threshold is
determined from the pixel value distribution in the sky region.  We
measure the mean pixel value in circular regions starting with a
radius of $2\farcs5$ and increasing in steps of $2\farcs5$, out to a
radius $\sim\!120\arcsec$.  The value of $f_0$ is determined by
averaging pixel values between radii of $40\arcsec$ and $45\arcsec$.
These radii were chosen based on a flattening of the 24~\micron\
profile in this region.  At radii between $45\arcsec$ and
$\sim\!55\arcsec$ the profile continues to rise, likely due to
contamination from other nearby sources.  Using larger values of $f_0$
would artificially increase $\tau$.  This choice of $f_0$ is further
justified by the 8.0 and 5.8~\micron\ profiles: they show an increase
in the mean pixel value well past $40\arcsec$ but a decreasing value
approaching the inner estimate at radii larger than $\sim\!80\arcsec$,
and by a radius of $\sim\!100\arcsec$ the profiles match the inner
values.  This increase in the profile values at intermediate radii may
be caused by limb brightening of the core as it is irradiated by the
interstellar radiation field.  If the emission raises the observed
shadow flux then it will cause an underestimate of $\tau$ at all
wavelengths; however, if it predominantly increases the unobscured
flux levels outside the shadow $\tau$ will be over--estimated.  We do
not attempt to correct for this effect.  We calculate optical depth
limits of $\tau_{24} = 2.1, \tau_{8.0} = 3.2$ and $\tau_{5.8} = 2.2$
through the center of the profiles.  The implied total masses at these
wavelengths is $\sim\!1.2$~\msun\ for all three profiles, assuming
\citet{ossenkopf94} model dust opacities.  The discrepancy between
this mass and the 70~\micron\ mass (between 1.6 and 5.1~\msun) most
likely results from the high optical depth in the core.  If the core
is dense enough to be observed in absorption at 70~\micron\ then it is
certainly opaque at shorter wavelengths; the $A_V$ profile and mass
estimates at shorter wavelengths will be lower limits.  In the CB190
case the core was observed in absorption at 24~\micron\ but not at
70~\micron, and the maximum measured extinction value was $A_V \sim
30$~mag ($\tau_{24} = 1.4$), much smaller than the values measured
here \citep{stutz07}.  Therefore we reject all 24, 8.0, and
5.8~\micron\ $\tau > 1.0$ points and we scale the profiles to match
the 70~\micron\ extinction values at the outer two 70~\micron\ radii.
In Figure~\ref{fig:av} we plot the scaled profiles; the shapes of the
profiles match well.

\section{Discussion}

Bonnor-Ebert models are often used to characterize starless cores;
these models generally fit the observed density profiles with
reasonable fidelity \citep[e.g.,][]{kandori05,shirley05,stutz07}.
Bonnor-Ebert models describe a self--gravitating pressure confined
isothermal sphere in hydrostatic equilibrium.  The density profiles
are described by three parameters: temperature, central density
$\rho_{\rm c}$ (or equivalently, $\xi_{\rm max}$), and the outer
radius of the cloud.  A model will be unstable to collapse if
$\xi_{\rm max} > 6.5$.  We fit said models to the extinction profiles
discussed above.  We use a 3D grid of models in temperature, radius,
and $\xi_{\rm max}$; see \citet{stutz07} for a more detailed
discussion of the models and the fitting method.  One strong
constraint in interpreting the best--fit models is that \cbo\ is dark
at $\lambda \lesssim 70$~\micron, and therefore the temperature cannot
be greater than $\sim\! 12$~K, in agreement with other core
temperature measurements
\citep[e.g.,][]{lemme96,bacmann02,hotzel02,kirk07}.  The best--fit
models for the upper and lower bounds on the 70~\micron\ profile are
shown in Figure~\ref{fig:av} as grey dashed lines.  The 70~\micron\
profile has a best--fit temperature of $T = 56-130$~K, for the upper
and lower bounds, respectively; these temperatures are in clear
contradiction with the data.  The best--fit outer radius is
$27\arcsec$, for both the upper and the lower limit, and the best--fit
$\xi_{\rm max}$ = 4.5 and 3.0, for the upper and lower limit
respectively.  As the assumed distance is increased (in this case to
270~pc) the best--fit models become even more implausible, with
temperatures always $>70$~K.  At a distance of 100~pc the best--fit
temperatures are $T = 26-90$~K, still in contradiction with the data.
These high temperatures indicate that \cbo\ requires additional
support (turbulence and magnetic fields) beyond thermal to prevent
collapse.

\citet{crapsi05} observe N$_2$D$^+$ and N$_2$H$^+$ line widths
$\sim\!0.3 - 0.4$~km~s$^{-1}$ and calculate excitation temperatures
$T_{\rm ex} \sim\!4.5 - 5$~K.  The particle of mean molecular mass has
velocity FWHM $\sim\!0.4$~km~s$^{-1}$ for temperature 8 K, indicating
that the turbulent energy contribution to the support in \cbo\ is only
comparable to the thermal contribution \citep{hotzel02}.

Magnetic fields are a possible source of support in starless cores.
Inferred magnetic field strengths in cores range from, e.g.,
$\lesssim15~\mu$G to $160~\mu$G \citep{bergin07}.  For $\theta =
30\arcsec$, a distance of 200~pc, and $B = 160$~$\mu$G, the mass that
can be supported is $M_{\rm B} = 1$~\msun\ \citep{stahler05}.
Furthermore, from equation 1 and the \citet{stahler05} analysis, the
condition of stability requires a minimum magnetic field strength of:
$$
B = 861~\mu {\rm G}  
\left( \frac{1.58 \times 10^2~{\rm cm}^2~{\rm gm}^{-1}}
     {\kappa_{\rm abs,70}}\right) 
\left( \frac{M_{\rm B}}{M_{70}} \right) 
\left( \frac{\theta}{30\arcsec} \right)^{-2},
$$ where $\theta$ is the angular radius of the core.  The mean column
density of the 70~\micron\ profile is $\langle N({\rm H_2})\rangle
\simeq 1.1\times10^{23}$~cm$^{-2}$; given this column, the correlation
presented by \citet{basu04} implies a line-of-sight magnetic field of
$B_{\rm los} = 250$~$\mu$G; we estimate the total field as $B_{\rm
tot} = \sqrt 3 \times B_{los} = 430$~$\mu$G, although this relation
has large scatter.

Although the field required to support \cbo\ is larger than expected,
our understanding of the limits on magnetic field strengths does not
allow for a strong constraint.  In addition to requiring a field
exceeding the likely range of values, supporting the globule entirely
with magnetic pressure has other consequences. One might expect
asymmetries to develop, given the asymmetric pressure provided by
magnetism.  If such structures can be avoided \citep{shu87}, another
issue is that the field should leak out of the globule by ambipolar
diffusion, on timescales $\sim\!3\times10^6$~yr
\citep{stahler05}. Thus, either the field is already too small to
prevent the collapse of the globule, or the collapse will only be
delayed for a time.

If infall is occurring in \cbo, the observed line width is very
narrow.  \citet{crapsi05} detect a double-peaked N$_2$H$^+$ $(1-0)$
line and \citet{lee99} detect a double-peaked CS~$(2-1)$ line, both
measurements with roughly equal brightness in the blue and red peaks.
\citet{lee99} state that because other thin tracers show single
Gaussian components, the observed double peak is likely due to self
absorption at high optical depth; these authors also note that \cbo\
is one of two sources out of 163 spectra that show small wing
components.  \citet{lee04} classify \cbo\ as a possible infall
candidate based on their observation of CS~$(3-2)$, which has a double
peaked profile with more emission in the blue peak.  \citet{sohn07}
also measure a double peaked profile in HCN $(1-0)$; they note that
this source is one of two (out of 85 cores surveyed) that has some
line strength anomalies in the three hyperfine HCN lines.  Recently,
\citet{caselli08} have found that \cbo\ has a large amount of
ortho-H$_2$D$^+$(1$_{1,0}$--1$_{1,1}$) emission, which they find to be
correlated with increased central concentration, CO depletion, and
deuteration, providing further evidence that \cbo\ is very evolved.

\section{Conclusions}

We measure the density profile of \cbo\ from the 70~\micron\
absorption feature observed with the {\it Spitzer} Telescope.  We
conclude that:

\noindent 1. The strong absorption even at 70~\micron\ indicates that
\cbo\ is exceptionally dense.

\noindent 2. At the same time, the absorbing cloud is very compact and
has a very steep density gradient (as seen most clearly at
24~\micron).

\noindent 3. The temperature can be no more than 12~K.

\noindent 4. As a result of these extreme characteristics, thermal
pressure fails, by a wide margin, to support the globule; turbulent
support is also inadequate by a wide margin.

\noindent 5. The magnetic field required to make up the rest of the
required pressure for support would be surprisingly large; even if
present, such a large field will leak out of the globule through
ambipolar diffusion.

\noindent 6. Therefore, L429 is either already undergoing collapse or
is approaching an unstable, near-collapse state.

\

We thank the referee for their helpful comments.  Support for this
work was provided through NASA contracts issued by Caltech/JPL to the
University of Arizona (1255094) and to SAO (1279198 and 1288806).
This work was also supported by the National Science Foundation grant
AST-0708131 to The University of Arizona.

\end{document}